# EFFECT OF GAMMA IRRADIATION ON THE STRUCTURE AND VALENCE STATE OF Nd IN PHOSPHATE GLASS

V. N. Rai[1], B. N. Raja Sekhar[2], D. M. Phase[3], and S. K. Deb[1]

[1]Indus Synchrotron Utilization Division, Raja Ramanna Centre for Advanced Technology

Indore-452013, (India)

[2] Atomic and Molecular Physics Division, Bhabha Atomic Research Centre

Mumbai- 400085, (India)

[3]UGC-DAE Consortium for Scientific Research, University Campus, Khandwa Road,

Indore, (India)

Address for Correspondence

E-mail: vnrai@rrcat.gov.in

Phone: +91-731-2488142

# ABSTRACT

Fourier transform infrared (FT-IR) spectra and X-ray photoelectron spectra (XPS) of Nd doped phosphate glasses have been studied before and after gamma irradiation in order to find the behavior of chemical bonds, which decide the structural changes in the glass samples. IR absorption spectra of these glasses are found dominated mainly by the characteristics phosphate groups, water (OH) present in the glass network as well as on the composition of glass matrix. The effects of gamma irradiation are observed in the form of bond breaking and possible re-arrangement of the bonding in the glass. Energy dispersive X-ray spectroscopy (EDX) and XPS measurements show changes in the relative concentration of elements; particularly decrease in the concentration of oxygen in the glass samples after $\gamma$-irradiation, a possible source of oxygen vacancies. The decrease in the asymmetry in O 1s spectra after gamma irradiation indicates towards decrease in the concentration of bridging oxygen as a result of P-O-P bond breaking. Asymmetric profile of Nd $3d_{5/2}$ peak after gamma irradiation is found to be due to conversion of $Nd^{3+}$ to $Nd^{2+}$ in the glass matrix.

## 1.  INTRODUCTION

Recent technological applications have generated more interest in the studies of glasses [1-9]. The doping of rare earth ions has been extensively investigated in various glasses and crystals, because they play an important role in the development of many optoelectronic devices such as lasers, light converters, sensors, high density memories, optical fibers and laser amplifiers [10-13]. The spectroscopic and lasing properties of rare earth ions are strongly affected by the local structures at the rare earth sites and the distribution of the rare earth doped ions in the glass matrix [14-15]. The local structural properties are expressed by the type and arrangement of the ligands surrounding the rare earth ions. The study of doped glasses is useful for designing laser glasses and other optical components. Various types of glasses such as silicates, phosphate, borates, fluorides and telluride have been used as matrix for doping trivalent rare earth ions to produce active optical devices including lasers, infrared to visible up-conversion phosphors, etc [1-13]. Silicate glasses are among the most studied one after its use in the fabrication of first glass laser [16]. Several compositions and different preparation techniques have been developed in order to improve and better understand the laser action and the glass properties [17-18]. As a result of these investigations Nd-phosphate glasses have been widely used as bulk laser materials [19-20]. Among oxide glasses, phosphate glasses are particularly attractive host because they can accommodate large concentrations of active ions without losing the useful properties of the material. Phosphate glasses are relatively easy to prepare and offer an important range of compositional possibilities, which facilitate tailoring of the physical, chemical and optical properties of interest for specific technological applications [21-26].

It has been established that most of the glasses produce various interesting changes in their properties, when irradiated by high energy particles or radiation such as UV, $\gamma$ rays and neutrons [27-36]. The irradiation effect produces various changes in the optical properties of glasses in the form of generation of some new absorption bands. Recently a series of studies has been reported on $\gamma$-irradiation effect on different types of glasses such as soda lime phosphate, Cabal, bioglass, borosilicate, lithium borate and phosphate glasses [37-40], where intrinsic defects are found due to presence of dopants or impurity in the glass. Some of the dopants such as transition metal ions or lanthanide ions dopants capture negatively charged electrons or positively charged holes creating defects in the form of a change in their valence state through photo chemical reaction during the exposure to successive gamma irradiation. Zang et al. [41] reported that $Yb^{3+}$ changes to $Yb^{2+}$ after $\gamma$ irradiation in YAG crystal, which recovers after annealing. Similar changes in valence state of $Nd^{3+}$ to $Nd^{2+}$ have been reported by Rai et al. [28, 36, 42] in neutron irradiated $Nd_2O_3$ powder and gamma irradiated Nd doped phosphate glasses. Generation of other types of defects such as oxygen vacancies and color centers have also been reported earlier [37-39]. Recently ElBatal et al. [40] studied the $NdF_3$ doped borophosphate glass before and after gamma irradiation. They reported either very less or no generation of defects after irradiation, which may be due to better stability of the mixed phosphate and borate units causing the compactness and resistance towards gamma irradiation. Normally most of the above reports are made on the basis of UV and visible spectroscopy. The presence of different dopant, its chemical effects and the role of network modifiers in the phosphate glasses are quite complex. Numerous efforts have been made to help determine the structural role of network modifier groups and defects in glasses using optical absorption, FTIR, Raman, NMR, XPS spectroscopy and X-ray diffraction etc. FTIR [33-35, 45] provides qualitative probing of the local structure of phosphate glasses, whereas XPS [43-46] is particularly useful for providing access to the core electrons energy of the various atoms. The glass structures and other physical properties of the glasses were also explained in terms of valance state of the dopant along with its identification using XPS. Recently Khattak et al. [47] have reported the effect of laser irradiation on the local glass structure as well on the valence state of the vanadium ions in phosphate glass containing $V_2O_5$, using X-ray photoelectron spectroscopy (XPS). This shows that XPS can be helpful in identifying the radiation induced change in the valence state of dopants, particularly $Nd^{3+}$ to $Nd^{2+}$ in Nd doped phosphate glasses. Similarly it is also possible to distinguish the behavior of bridging and non-bridging oxygen atoms in glasses under the effect of different structural units and gamma irradiation.

This paper presents the study of Nd doped phosphate glass before and after gamma irradiation using FTIR and XPS. It is useful in order to better understand, how compositions of the glass as well as gamma irradiation affect short range structure of the glass along with certain other properties. The generation of different types of defects such as oxygen vacancies and change in the valence state of Nd ($Nd^{3+}$ to $Nd^{2+}$) in the phosphate glass after gamma irradiation are also discussed.

## 2. Experimental

### 2.1 Preparation of the glasses

The Nd doped phosphate glass samples (Obtained from CGCRI, Kolkata) were prepared using different composition of $P_2O_5$, $K_2O$, $BaO$, $Al_2O_3$, $AlF_3$, SrO and $Nd_2O_3$ as base materials. Different combination of base materials were used for getting four types of glass samples such as, sample # 1 ($P_2O_5$, $K_2O$, BaO, $Al_2O_3$, $Nd_2O_3$), sample # 2 ($P_2O_5$, $K_2O$, BaO, $Al_2O_3$, $Nd_2O_3$), sample # 3 ($P_2O_5$, $K_2O$, BaO, $Al_2O_3$, $Nd_2O_3$, $AlF_3$) and sample #4 ($P_2O_5$, $K_2O$, SrO, $Al_2O_3$, $Nd_2O_3$). The weight percentage of each oxide taken for making sample # 1 to sample # 4 is given in table -1. Melt quenching technique was used to make these glasses, where reagents were thoroughly mixed in an agate mortar and placed in a platinum crucible for melting it in an electric furnace at $1095^0$ C for 1h 40 m. The melt was then poured onto a preheated brass plate and annealed at $365^0$ C for 18 h. Finally, the samples were polished to obtain smooth, transparent and uniform surface slab of 5 mm thickness for optical measurement. The elemental compositions of each element (atomic %) in the glasses were measured after glass formation using energy dispersive X-ray spectra (EDX). Spectra of each sample were recorded before and after γ-irradiation using Bruker X-Flash SDD EDS detector, 129 eV in order to obtain the relative concentration of different elements present in the glass samples. Due to variation in concentration of elements from one place to other in the same sample, final data was obtained after averaging the data recorded at three random locations. Tables 2 to 4 show the average atomic % of the important elements present in the glass samples before and after gamma irradiation.

### 2.2 γ- Irradiation of samples

Few small pieces of glass samples were irradiated at room temperature using $^{60}$Co (2490 Ci, Gamma chamber 900) source of gamma radiation having dose rate of 2 kGy/h. Samples were irradiated for radiation doses varying from 10 to 500 kGy.

### 2.3 FT- IR spectroscopy

FT-IR spectra of the glass samples were recorded before and after γ-irradiation using KBr pellet technique. For this purpose glass samples were ground in a clean mortor to a fine powder. The glass sample powder (1 % by wt.) was mixed with pure KBr powder. The mixture was then pressed with a pressure of 2 ton per square inch to yield a transparent pellet of approximate thickness ~ 0.1 mm suitable for mounting in the FTIR spectrophotometer (Brucker Model 80V). FT-IR spectra of all the glass samples were recorded before and after irradiation in the wave number range from 400 – 4000 cm$^{-1}$ at room temperature.

The values of differential absorption or additional absorption (AA) spectra due to the irradiation are calculated using the formula [48] as

$$\Delta k = \frac{1}{d} \ln \frac{T_1}{T_2} \qquad \text{------------------------------------- (1)}$$

Where d is the sample thickness and $T_1$ and $T_2$ are the values of transmission intensity at particular wavelength obtained from the spectra of Nd doped phosphate glass recorded before and after γ-irradiation respectively. In fact $\Delta k$ is just a difference between linear absorption coefficient after and before irradiation, where bulk contribution cancels out and resulting spectrum remains mainly due to induced defects.

### 2.4 XPS Spectrum

The XPS measurements were carried out using a spectrometer equipped with Al $K_\alpha$ X-ray source. X-Ray photoelectron spectra mainly from O1s and Nd 3d core level were recorded using a computer controlled data collection system. X-Ray photoemission measurements were performed using a non-monochromatic Al $K_\alpha$ radiation (hν = 1486.6 eV). The recorded spectra have been reported without any correction for charging effect. In this experiment XPS results have been used to find qualitative informations to support the FT-IR data from glass samples.

## 3. RESULTS AND DISCUSSION

### 3.1 Infrared spectrum of Nd phosphate glass

FT-IR spectra of Nd doped phosphate glasses are shown in Fig.-1, whereas Table-5 shows the assignment of infrared absorption bands obtained in the present investigation along with reported assignments by others for different matrix of phosphate glasses. FT-IR spectra of samples in the frequency range 500 to 1400 $cm^{-1}$ show a number of strong characteristics bands of phosphate glass. There is no significant difference in the line shapes of the spectra, because all the samples contain $\geq$ 50 mol. % $P_2O_5$, which is as expected. These spectra consist of two main sets of IR absorption bands in the low and high frequency regions due to phosphate group and OH molecule respectively. The assignment of IR spectra shows the presence of a broad band peak around 3440 $cm^{-1}$, which is due to the presence of $H_2O$ molecules in the glass structure. It is possible due to the absorption of atmospheric moisture by the sample or by the pellet resulting in the appearance of IR band belonging to $H_2O$ molecules, because the sample as such does not contain $H_2O$ as unit in the network. The three bands located in the range 2346 – 2926 $cm^{-1}$ are relatively weak and can be assigned to the stretching vibrations of P-O-H group in different structural sites. A weak band at 1640 $cm^{-1}$ is assigned due to P-O-H bridge. These groups form the strongest hydrogen bonding with the non bridging oxygen. The shoulder observed at 1390 $cm^{-1}$ is due to the characteristic stretching mode of P=O bond. Several authors have reported that the band corresponding to the stretching vibration of doubly bonded oxygen could be found in the frequency range 1230 – 1390 $cm^{-1}$. The band at about 1287 $cm^{-1}$ is assigned to the asymmetrical stretching vibration of O-P-O and/or P=O groups, $\nu_{as}$(O-P-O)/ $\nu_{as}$ (P=O), while the medium broad band at 1103 $cm^{-1}$ is related to the symmetric stretching vibration of that groups, $\nu_s$(O-P-O). The 1103 $cm^{-1}$ band is the dominant feature of the spectra and has been assigned to P-O$^-$ groups. The phosphorous non-bridging oxygen portion of $PO_4$ tetrahedra in a chain structure has been referred to as a P-O$^-$ unit. The asymmetric stretching vibration of the metaphosphate group, $\nu_{as}$ ($PO_3$) reported previously in the range 1080-1120 $cm^{-1}$ is observed in the range 1027 – 1161 $cm^{-1}$. This indicates that the bands due to $\nu_{as}$ ($PO_3$) can interfere with the band at 1103 $cm^{-1}$. The broadening observed in the region 900-1105 may be due to the interference of $\nu_s$ ($PO_3$) with the spectral range 1000-1060 $cm^{-1}$. The shoulder around 903 $cm^{-1}$ is assigned due to the asymmetric stretching vibration of P-O-P linkages, $\nu_{as}$(P-O-P), while the relatively weak band around 764 $cm^{-1}$ is due to the symmetric stretching vibration of that linkage, $\nu_s$(P-O-P). The absorption band at about 525-539 $cm^{-1}$ may be assigned to the harmonics of P-O-P bending vibration as well as due to deformation mode of PO$^-$ group. The low frequency absorption band around 475 cm$^{-1}$ is either due to bending vibration of O-P-O or harmonics of O=P=O linkages.

A comparison of spectra in Fig.-1 shows that sample # 1 and # 2 shows nearly similar IR transmission (absorption) spectra, which may be due to similar composition of the glass matrix except different amount of Nd, which seems to be having less or negligible effect on IR absorption spectra. Spectra of sample # 3 and # 4 show less overall transmission (increased absorption), which may be due to the effect of extra addition of AlF$_3$ and SrO in the samples respectively. Similar increase in overall IR absorption by addition of extra network modifier has been reported by ElBatal et al. [30, 33]. The addition of AlF$_3$ in the network shows decrease in the amplitude of all the bands having P-O-H bonding indicating decrease in the presence of moisture. This observation is in agreement and in support of reported results [4, 23] that addition of fluoride content in the network increases the resistance to water and hence decreases the probability of non-radiative relaxation from the emitting levels, which seems to be due to changes in local environment of Nd$^{3+}$ ions in the network. The addition of SrO in sample # 4 produces various new peaks in the band 900-1165 cm$^{-1}$ ( near P-O-P band) and in the band 1287 – 1750 cm$^{-1}$ (near P=O band). The intensity of bands due to P-O-P and P=O linkages decreases. It seems that both types of bonds are breaking resulting in some new linkages possibly as P-O-Sr. The new linkages may be producing change in the structure of band 900-1165 cm$^{-1}$. Although there is no sufficient evidence for the existence of band characteristics of P-O-Nd, P-O-Sr or P-O-M, where M is other cations present in the glass matrix. Here it is presumed that the stretching band corresponding to P-O-Nd, P-O-Sr and P-O-M linkages might locate between 900-1165 cm$^{-1}$, but not observable due to overlapping with other bands in the range. Similar idea has been proposed by Xu and Day [48] for Sn-P-O-F glasses. Such overlapping of absorption bands gives rise to difficulties in analysis

The absorption band at 1260 cm$^{-1}$ which corresponds to the PO$^-$ stretching vibration tends to move to higher frequency area with other oxide substitutions. The intensity of this band and other bands in spectra do not significantly alter with composition, while the position of each band does shift slightly with change in oxide composition [45]. The absorption band at about 880-900 cm$^{-1}$, which is the characteristics band for P-O-P stretching, moves to higher frequency by substitution of other oxides. The effect of this substitution is an increase in cation oxygen attraction as a result of its comparatively high field strength.

### 3.2 Structural information

The analysis of IR absorption spectra of phosphate glasses in section 3.1 has been made after taking into consideration some of important criteria, which are discussed below. The network structure of phosphate glasses is accepted to contain a polymeric arrangement of phosphate groups with variable lengths depending upon the other partners [49-50]. This network is dominated by linkages between $PO_4$ tetrahedra. The $(PO_4)^{3-}$ tetrahedra that make up the phosphorous-oxygen network in a phosphate glass are grouped into one of four categories as $Q^3$, $Q^2$, $Q^1$ and $Q^0$ based on the number of bridging oxygen [51]. The superscript in this $Q^n$ notation refers to the number of bridging oxygen in $(PO_4)$ unit. Based on the composition of the oxygen to phosphorus ratio, phosphate glasses are classified as ultraphosphates ([O]: [P] < 3), mataphosphates ([O]: [P] =3) and polyphosphates ([O]: [P] > 3. Considering this classification we found that our samples #1 to # 4 come under category of polyphosphate glass as [O]: [P] ratios for sample #1 to # 4 are 5.90, 8.02, 8.11 and 5.7 respectively. It is accepted that the structure of polyphosphates are dominated by $Q^2$ and $Q^1$ units. Normally an addition of alkali or alkaline earth oxides to $P_2O_5$ glass results in conversion of three dimensional networks to linear phosphate chains [52-53]. This linear chain structure results in the cleavage of P-O-P linkages and the creation of non-bridging oxygen in the glass. However, the behavior of network former is thought to be quite different from that of a network modifier. It has been reported that the combination of trivalent Al and pentavalent P in tetrahedral coordination allows for a local structure upon addition of $Al_2O_3$ to phosphate glasses [54]. It has been suggested that such structure of AlPO4 is formed by opening up of the P=O bond and forming P-O-Al linkages. Similarly Selvaraj and Rao [55] has proposed a model for the role of PbO in phosphate glasses that P-O-Pb linkages may be formed by opening up P=O bonds of $PO_4$ tetrahedra. Similar view point was also used to interpret the formation of covalent P-O-Cu linkages in copper phosphate glasses [46]. However, further investigation is needed to find conclusive evidence supporting these assumptions.

However, most of the research on vibrational spectroscopy of phosphate glasses treats information on their structure obtained in terms of the oxygen bridges and terminal groups rather than in terms of polyhedral or $Q^n$ units [56-57]. Notably, it is the $(PO_2)^-$, $(PO_3)^{2-}$ and $(PO_4)^{3-}$ groups rather than separate P=O and P-O groups. Finally, realization and interpretation of infrared spectra have been carried out according to the concept adapted by Tatre [58-59] and Condrate [60] about the independent vibrations of different groups in the glass. Similar procedure has been used successfully by many authors [33-35, 37-38]. It has been reported [18] that due to structural disorder or aperiodic arrangement a complete vibrational analysis in the glass is not possible from an ab initio point of view and an analysis is done primarily by comparing the spectra of those glasses with crystalline phosphate for which the structures are known.

Careful observation of present results and its comparison with other reported results indicate that most of the IR bands observed in the spectra correspond to the phosphate groups and moisture. However, the small changes in the amplitude and frequencies of the bands may be due to different linkages (groups) present due to changes in the matrix composition. No band was observed for Nd-O-P bonding probably due to low concentration of Nd. The appearance of several absorption bands associated with the P-O-P linkages may be the result of formation of metaphosphate chain with different lengths.

### 3.3  IR spectrum of gamma irradiated phosphate glass

Fig.-2 shows the IR spectra of sample # 1 before and after gamma irradiation (10 & 500 kGy). It seems that nature of the spectra and its peak remains similar after $\gamma$ irradiation except a change in the intensities of the absorption peaks along with generation of some new peaks. It shows that after irradiation all the peaks due to phosphate group (400 – 1500 cm$^{-1}$) show decrease in transmission, where as in the spectral range 1500 – 4000 cm$^{-1}$ transmission increases for 10 kGy and decreases for 500 kGy irradiation in comparison to unirradiated sample. Main observation after 10 kGy irradiation is the generation of some new peaks in the 800-1017 cm$^{-1}$ range dominated by P-O-P linkages. This clearly indicates the breaking of these P-O-P linkages, which produces more non-bridging oxygen. This is associated with an increase in the intensity

of band due to P=O bond at 1187 cm$^{-1}$ (non-bridging oxygen), which confirms the idea of breaking P-O-P linkages in the glass. However further increase in dose of irradiation to 500 kGy shows decrease/absence of new peaks. The transmission intensity also recovers near to unirradiated sample. This may be possible after reorganization of the bonds in the glass sample. In the case of sample # 3, IR transmission increases (absorption decreases) in all the frequency range (400-4000 cm$^{-1}$) after 10 kGy irradiation. Similar changes in IR spectra have been reported earlier [33]. The γ-irradiation changes the IR transmission spectra probably due to breaking of bonds in the glass network. The differential spectra of sample # 1 and # 3 irradiated for 10 kGy were obtained using eq.-1 and are shown in Fig.-3 in order to find the changes in IR spectra after same dose of γ irradiation. A comparison shows that peaks in the difference spectra are similar for both the samples except some changes such as their amplitudes, broadening in peaks and back ground absorption along with development of some new peaks. The amplitudes of peaks due to phosphate group (< 1500 cm$^{-1}$) in sample # 3 show an increase along with broadening except the peak at 475 cm$^{-1}$. The band due to P-O-H bond near 2900 cm$^{-1}$ has decreased amplitude in sample # 3, which may be due to the presence of fluorine in the sample as discussed in previous section. Development of new peaks after γ irradiation indicates towards formation of new bonds in the glass network. However drastic change in the back ground absorption in sample # 3 seems to be occurring due to addition of AlF$_3$ in the glass matrix. Finally these results indicate that addition of AlF$_3$ in the matrix makes sample # 3 soft for γ irradiation in comparison to sample # 1.

### 3.4 *Effect of gamma irradiation on glasses*

The changes observed in IR spectra of glass samples after γ-irradiation indicates that bond breaking as well as rearrangement of bonds may be taking place during γ-irradiation, which is dependent on doses of irradiation and composition of glass material. Hobbs et al [61] reported that irradiation induced defects effectively break the connectivity of the network, whereas accumulation of such broken linkages results in local structural collapse as well as stochastic rebinding. Piao et al [62-63] describes that production of electron hole pairs during irradiation provides another paths for bond rearrangement reducing the constraints on structural relaxation. The relaxation process release some of the excess energy stored in the structure resulting in change in the bond angle. Similar observations have also been reported by Sharma et al. [29].

## 3.5.    *Effect of γ-irradiation on XPS of phosphate glass*

### 3.5.1 *XPS spectra of Nd 3d*

X- Ray photoelectron spectra (XPS) of Nd doped phosphate glasses were recorded before and after γ-irradiation using Al $K_\alpha$ line emission. X-ray was focused on the surface of the sample. Core level of Nd 3d spectrum was studied. Fig.-4   shows the XPS spectra of glass sample # 1 before and after 10 & 500 kGy γ-irradiation. The inbox shows two broad peaks at 982.5 and 1003.7 eV due to $3d_{5/2}$ and $3d_{3/2}$ (Nd 3d spectra). These peaks become sharper after gamma irradiation and shifts to higher kinetic energy. Observation of sharpening and shift in the main peak towards higher energy side may be due to decrease in oxygen content in the glass sample after γ-irradiation, as is observed in EDX data, which clearly indicates that atomic % of oxygen content in sample # 1 decreases drastically (Table-2) after γ-irradiation (500 kGy).   This decrease in the concentration of oxygen in EDX data is not observable for low dose of irradiation as 10 kGy. Fig.-8 shows that sample #1 & # 3 also have sharp Nd 3d peak due to $3d_{5/2}$ after γ irradiation of 10 kGy. It is observed that main peak of sample # 3 shifts to lower energy side and is broad in comparison to sample # 1, which may be due to the presence of $AlF_3$ and higher concentration of oxygen in sample # 3 network. Further the intensity of peak due to Nd $3d_{5/2}$ shows decrease after gamma irradiation. The intensity of this peak is comparatively high in the case of 500 kGy than 10kGy irradiation. This change in intensity of peak seems to be due to change in the relative concentration of elements after gamma irradiation. The diffusion of ions from surface to bulk and vice versa along with emission of oxygen after gamma irradiation may be one of the reasons for this change in intensity. This is similar as reorganization of bonds as discussed in IR observations. Another reason for decrease in the intensity of the peak Nd $3d_{5/2}$ may also be due to decrease in the number density of $Nd^{3+}$ due to change in the oxidation state from $Nd^{3+}$ to $Nd^{2+}$. These observations are in agreement with optical observations reported by Rai et al. [36, 42] on these samples. They have noted that difference optical absorption spectra (after and before gamma irradiation) show decrease in the intensity at the locations where $Nd^{3+}$ lines are observed. This decrease is found dependent on the doses of gamma irradiation. Similarly photoluminescence of $Nd^{3+}$ also decreases after gamma irradiation. These observations also indicated towards decrease in the number density of $Nd^{3+}$ in the glass after gamma irradiation. These changes were supposed to be due to change in the oxidation state of $Nd^{3+}$ to $Nd^{2+}$. The profile of the peak due to Nd $3d_{5/2}$ shows asymmetry towards lower energy side (Fig.-4 and 5) after gamma irradiation. This also indicates about the presence of another oxidation state of Nd after gamma irradiation, which may be $Nd^{2+}$. The correlation of above observations clearly indicates that gamma irradiation induces the change in the oxidation state of $Nd^{3+}$ to $Nd^{2+}$ in the glass sample.

XPS and EDX results indicate that concentration of oxygen decreases after $\gamma$-irradiation of glass samples. Similar variations in XPS of Nd containing alloy glass have been reported by Tanaka et al [64]. They have shown that the presence of higher concentration of oxygen in glass induces broadening in the peaks of $3d_{5/2}$ and $3d_{3/2}$ due to Auger O KLL and 3d satellite peaks. An O KLL peak represents the energy of the electrons ejected from the atoms due to the filling of the O 1s state (K shell) by an electron from the L shell coupled with the ejection of an electron from an L shell. This broadening due to presence of extra peaks is not observed in the sample, where oxygen content is less. The decrease in oxygen content is accompanied by a small shift in the main peak (Nd 3d) towards higher energy. Similar decrease in oxygen concentration after electron beam irradiation in glass samples has been reported by Puglisi et al [65]. They have used XPS to study the compositional changes in glass sample after electron beam irradiation and reported an out gassing of oxygen from the sample. It is well known that XPS is effective mainly on the surfaces as X-rays cannot penetrate the glass sample. Therefore, it cannot provide information about the bulk sample. As per the above discussion the decrease in intensity of Nd $3d_{5/2}$ may also be associated with diffusion of ions from surface to bulk. Recently Khattak et al. [47] have reported that concentration of Vanadium in Vanadium phosphate glass decreases after laser irradiation, which is supposed to be either due to evaporation of Vanadium from the surface of the glass or due to its diffusion from surface to bulk. The decrease in intensity may also be due to decrease in the number density of $Nd^{3+}$ due to change in the oxidation state from $Nd^{3+}$ to Nd$^{2+}$ after gamma irradiation as has been predicted earlier on the basis of optical observations [36,42].

### 3.5.2 XPS Oxygen Spectra O 1s

In most XPS studies of oxide glasses the O 1s spectra are more informative with respect to the structure of the glass than the cation spectra. Specifically, the binding energy of the O 1s electrons is a measure of the extent to which electrons are localized on the oxygen or in the inner-nuclear region, a direct consequence of the nature of the bonding between the oxygen and different cation. An asymmetry in the O1s core level peak indicates about the presence of two different types of oxygen sites in these glasses. Normally, the O 1s peaks for these glasses may arise from oxygen atom existing in some or all of the following structural bonds: P-O-P, P-O-Nd, Nd-O-Nd, P=O and P-O-M where M is other cations present in the glass matrix. Oxygen atoms that are more covalently bonded to glass former atom on both side are typically called bridging oxygen (P-O-P), while oxygen atoms that are more ionically bonded, atleast on one side or double bonded to a glass former atom are referred to as non-bridging oxygen atoms (P-O-Nd, Nd-O-Nd, P=O). More over since the bridging oxygen (BO) is covalently bonded to two glass former atoms, while the non- bridging oxygen (NBO) are ionically bonded only from one or both sides. The binding energy of NBO should be lower than that of the BO.

The O 1s spectra for sample # 1 before and after gamma irradiation (10 and 500 kGy) are shown in Fig.-6. A slight asymmetry is observed in the profile of O 1s peak, which is indicative of two different types of oxygen sites in these glasses. Asymmetry is observed towards higher energy side. Such O1s spectra have been studied in different types of glasses by many authors [44-47, 66]. They fitted O1s spectra by two Gaussian Lorenzian peaks in order to determine the peak position and the relative contribution of the different oxygen sites. Asymmetry in Fig.-6 indicates about the presence of less number of bridging oxygen in unirradiated glass samples. These profiles become nearly symmetric after gamma irradiation of 10 and 500 kGy as a result of decrease in number density of bridging oxygen and subsequent increase in density of non-bridging oxygen. Here exact reason for shift in the peaks is not known. However, it may be due to change in the concentration of different linkages in glass matrix along with space charge effect.

Similar asymmetry has been noted in the case of O 1s spectrum from sample # 3 as shown in Fig.-7. Here again asymmetry decreases after gamma irradiation of the sample indicating decrease in bridging oxygen. These observations are supporting the results obtained from FT-IR spectra indicating that gamma irradiation increases the breaking of P-O-P linkage and creating non-bridging oxygen in the form of P-O-Nd and/or P-O-M linkages.

## 4. CONCLUSION

FT-IR and X-ray photoelectron spectra of Nd doped phosphate glasses have been studied before and after $\gamma$-irradiation. The FT-IR absorption spectra of Nd doped phosphate glasses provide information about the main characteristics frequencies for phosphate bonds present in the glass network such as P=O, P-O-P, O-P-O and P-O-H. The presence of P-O-H bond indicates the hygroscopic nature of these phosphate glasses. Addition of $AlF_3$ in the glass matrix decreases the moisture content in the glass and increases the IR absorption. The decrease in oxygen content of glass sample after $\gamma$-irradiation as indicated by XPS and EDX measurements confirms bond breaking in the glass sample. This creates oxygen vacancy in the glass sample, which has been noticed as generation of defects in the optical observations. The decrease in asymmetry in O 1s peak after gamma irradiation indicates towards increase in the concentration of non-bridging oxygen after breaking of P-O-P linkage, which has bridging oxygen. The change in the elemental concentration in the glass after gamma irradiation is occurring probably as a result of diffusion of ions from the surface of the glass towards the bulk. An asymmetry observed in the XPS of Nd $3d_{5/2}$ peak indicates that gamma irradiation induces conversion of $Nd^{3+}$ to $Nd^{2+}$ in the glass, which is in agreement with the results obtained from optical absorption and photoluminescence measurements. Finally, the effect of gamma irradiation was observed in the form of change in the structural units of phosphate glass due to breaking and rearrangement of bonds in the glass network along with generation of various types of defects in the samples. All these changes are found dependent on the composition of the glasses as well as on the doses of irradiation.

**Acknowledgment**

Authors are grateful to R. J. Kshirsagar, P. Tiwari and S. Kher for their help and discussions during this work.

**FIGURE CAPTION**

1.  FT-IR spectra of Nd doped phosphate glasses, (1) Sample # 1, (2) Sample # 2, (3) Sample # 3, (4) Sample # 4

2.  FT-IR spectra of sample # 1 before and after $\gamma$-irradiation, (1) Pure sample # 1, (2) Sample # 1 irradiated for 10 kGy, (3) Sample # 1 irradiated for 500 kGy

3.  Differential infrared absorption spectra of glass samples after $\gamma$-irradiation for 10 kGy and obtained using eq.-1, (1) Sample # 1(10 kGy), (2) Sample # 3 (10 kGy)

4.  XPS spectra of sample # 1 before and after $\gamma$-irradiation, (1) Pure sample #1, (2) Sample # 1 irradiated at 500 kGy, (3) Sample # 1 irradiated for 10 kGy

5.  XPS spectra of glass samples after $\gamma$-irradiation of 10 kGy, (1) Sample # 1, (2) Sample # 3

6.  XPS spectra of oxygen O 1s, (1) Sample #1, (2) Sample # 1 irradiated at 10 kGy, (3) Sample # 1 irradiated at 500 kGy

7.  XPS spectra of oxygen O 1s, (1) Sample # 3, (2) Sample # 3 irradiated at 10 kGy

**Table - 1**  Material composition for making glass

| S. No. | Composition | Sam. #1 (wt. %) | Sam. #2 (wt. %) | Sam. #3 (wt. %) | Sam. #4 (wt. %) |
|---|---|---|---|---|---|
| 1. | $P_2O_5$ | 58.95 | 58.50 | 55.50 | 58.50 |
| 2. | $K_2O$ | 17.00 | 17.00 | 14.00 | 17.00 |
| 3. | BaO | 14.95 | 14.50 | 14.50 | ---- |
| 4. | $Al_2O_3$ | 9.00 | 9.00 | 9.00 | 9.00 |
| 5. | $Nd_2O_3$ | 0.10 | 1.00 | 1.00 | 1.00 |
| 6. | $AlF_3$ | ----- | ----- | 6.00 | ----- |
| **7.** | SrO | ----- | ----- | ----- | 14.50 |

**Table – 2**  Atomic % of elements present in the glass (EDX data)

| S. No. | Atomic Component | Sam. #1 (At. %) | Sam. #2 (At. %) | Sam. #3 (At. %) | Sam.#4 (At. %) |
|---|---|---|---|---|---|
| 1. | O | 77.84±6.35 | 82.32±5.50 | 81.06±5.60 | 77.87±7.47 |
| 2. | P | 13.20±0.60 | 10.27±0.43 | 9.99±0.43 | 13.64±0.73 |
| 3. | Ba | 1.94±0.45 | 1.32±0.26 | 1.77±0.37 | _________ |
| 4. | K | 3.96±0.25 | 3.58±0.16 | 4.73±0.20 | 3.19±0.20 |
| 5. | Al | 2.27±0.10 | 2.00±0.13 | 1.97±0.10 | 3.99±0.30 |
| 6. | Nd | 0.32±0.13 | 0.09±0.10 | 0.28±0.10 | 0.19±0.10 |

**Table – 3** Atomic % of elements present in Sample #1 before and after γ irradiation (EDX data)

| At. Comp. | O | P | Ba | K | Al | Nd |
|---|---|---|---|---|---|---|
| Av. At. % | 77.84 ±6.35 | 13.20 ±0.60 | 1.94 ±0.45 | 3.96 ±0.25 | 2.27 ±0.10 | 0.32 ±0.13 |
| Av. At. % (γ irradiated 10 kGy) | 75.45 ±4.40 | 12.86 ±0.50 | 2.38 ±0.30 | 5.27 ±0.2 | 3.08 ±0.10 | 0.30 ±0.10 |
| Av. At. Wt. % (γ irradiated 500 kGy) | 66.19 ±4.3 | 17.00 ±0.60 | 3.12 ±0.45 | 5.90 ±0.25 | 3.99 ±0.15 | 0.37 ±0.10 |

**Table-4**  Atomic % of elements present in the sample #3 before and after γ irradiation (EDX data)

| At. Component | O | P | Ba | K | Al | Nd |
|---|---|---|---|---|---|---|
| Av.At.% | 81.06 | 9.99 | 1.77 | 4.73 | 1.97 | 0.28 |
|  | ±5.60 | ±0.43 | ±0.37 | ±0.20 | ±0.10 | ±0.10 |
|  |  |  |  |  |  |  |
| Av. At. % | 82.83 | 7.21 | 0.49 | 0.99 | 3.85 | 0.34 |
| (γ irradiated 10 kGy) | ±8.40 | ±0.40 | ±0.20 | ±0.10 | ±0.25 | ±0.15 |

**Table-5**  Vibrational assignment of the FT-IR Spectra of glass samples

| S. No. | Vibrational Modes | Wave Numbaers (cm$^{-1}$) | | | | |
|---|---|---|---|---|---|---|
|  |  | Ref-34 | Ref-33 | Ref-35 | Ref-45 | Present |
| 1. | $H_2O$ or P-O-H vibration | 3433 3169 | 3420 | 3434 |  | 3440 |
| 2. | Stretching Vibration of P-O-H | 2700-2925 |  | 2923 2852 2419 |  | 2926 2855 2346 |
| 3. | P-O-H Bridge |  | 1640 | 1640 |  | 1640 |
| 4. | OH bending Vibration | 1601 | --- | --- |  | ---- |
| 5. | Stretching mode of P=O | 1390 1350 |  | 1360 |  | 1390 |
| 6. | Asymmetric stretching O-P-O |  |  |  | 1260 |  |
|  | $\nu_{as}$(O-P-O) | 1270 |  |  |  |  |
|  | $\nu_{as}$ (P=O) |  | 1270 | 1280 |  | 1287 |
| 7. | Asymmetric stretching of $PO_2$/$PO^-$ |  |  |  | 1190 | 1165 |
| 8. | Symmetric Stretching | 1105 |  |  | 1100 | 1100 |
|  | $\nu_s$(O-P-O)/$PO^-$ |  |  |  |  |  |
| 9. | $\nu_{as}$ ($PO_3$) | 1080-1120 | 1060 | 1096 1160 |  | 1027-1161 |
| 10. | $\nu_s$ ($PO_3$) | 900-1105 |  |  |  | 900-1105 |
| 11. | Asymmetric stretching P-O-P | 900 |  | 871 1011 | 880-900 | 903 |
|  | $\nu_{as}$ (P-O-P) |  |  |  |  |  |
| 12. | $\nu_s$ (P-O-P) | 764 | 725 | 727 769 | 740-780 | 708 740 764 |
| 13. | Harmonics of P-O-P bending Vib./Deformation mode of $PO^-$ | 536 | 500 | 523 | 530 | 525 539 |
| 14. | Bending Vibration (O-P-O) Harmonics O=P-O bending |  476 | 460-475 | 470-485 |  | 475 |

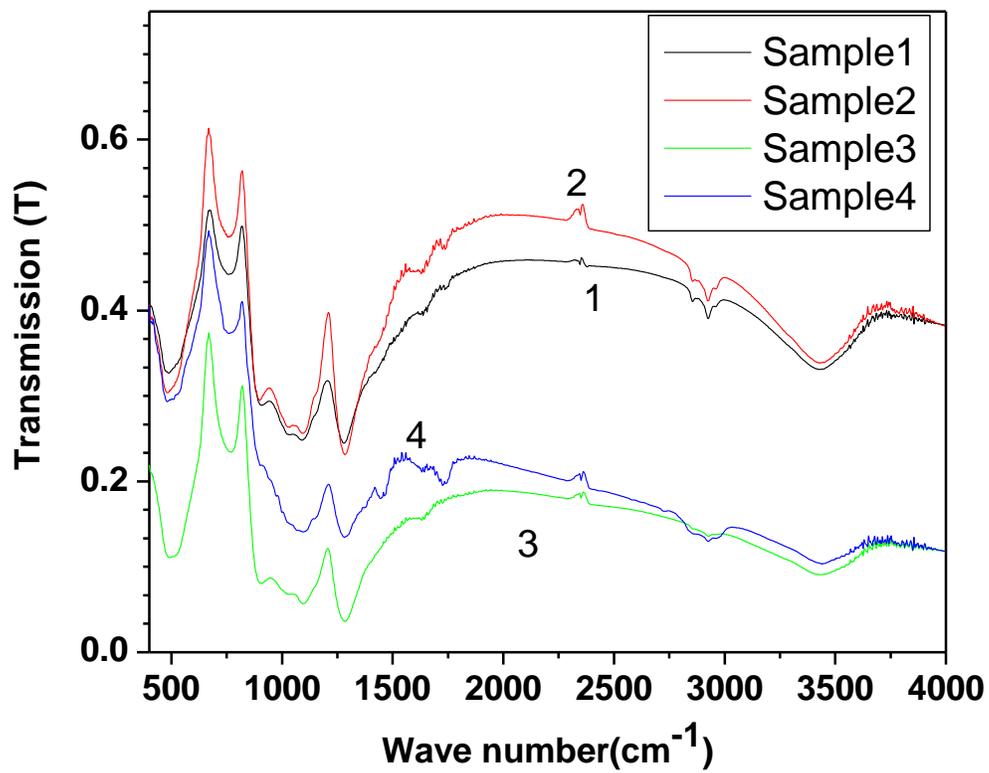

**Fig.-1**

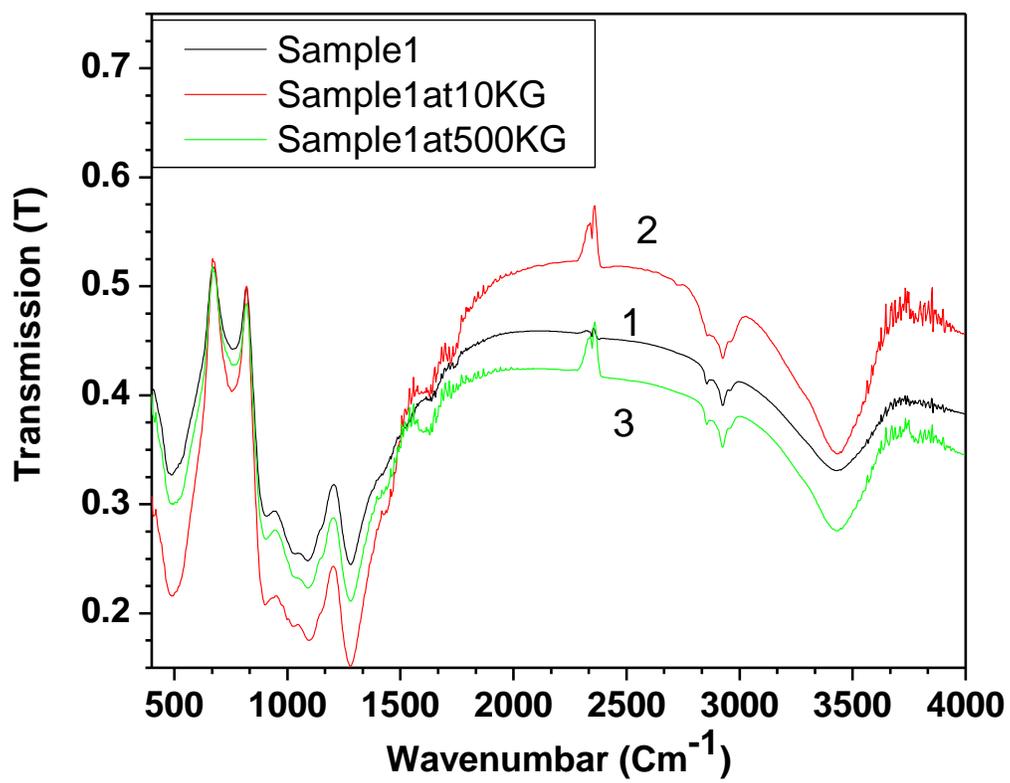

**Fig.-2**

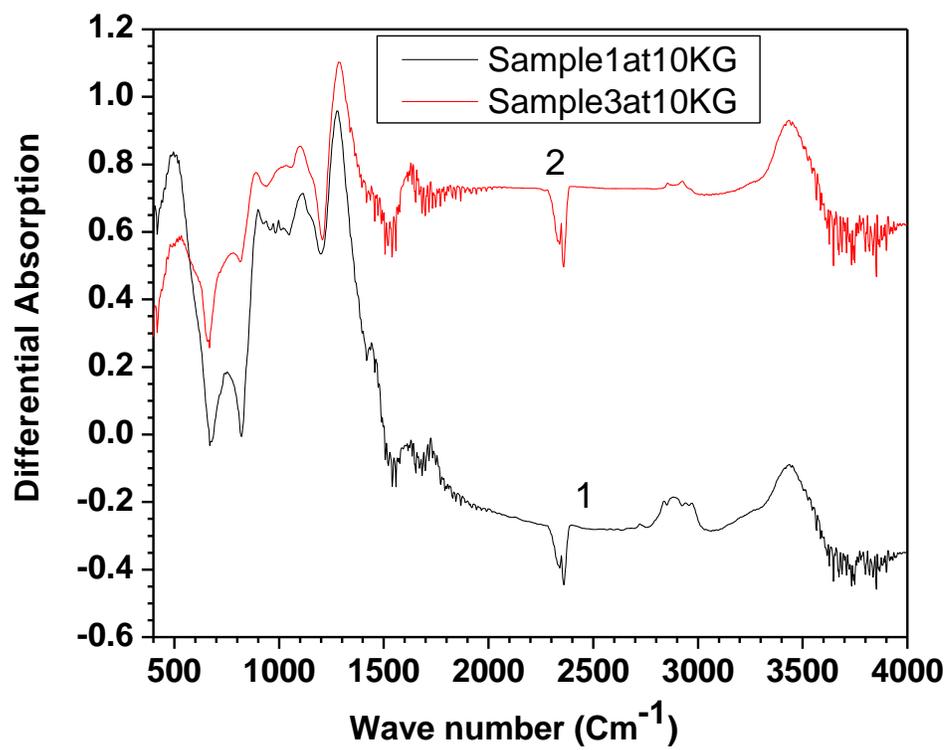

**Fig.-3**

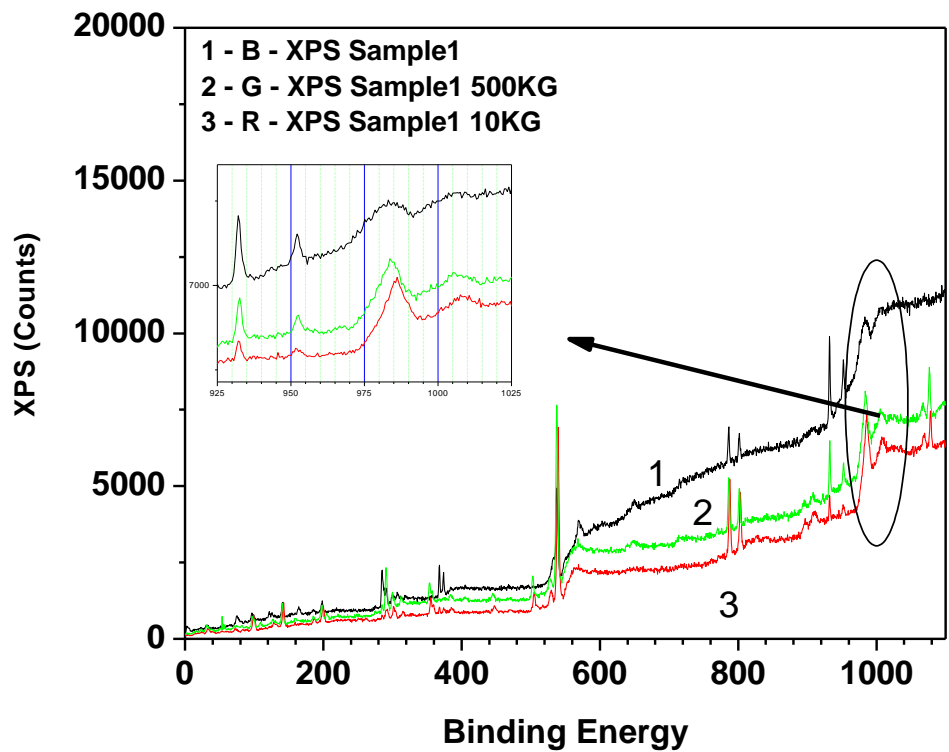

**Fig.-4**

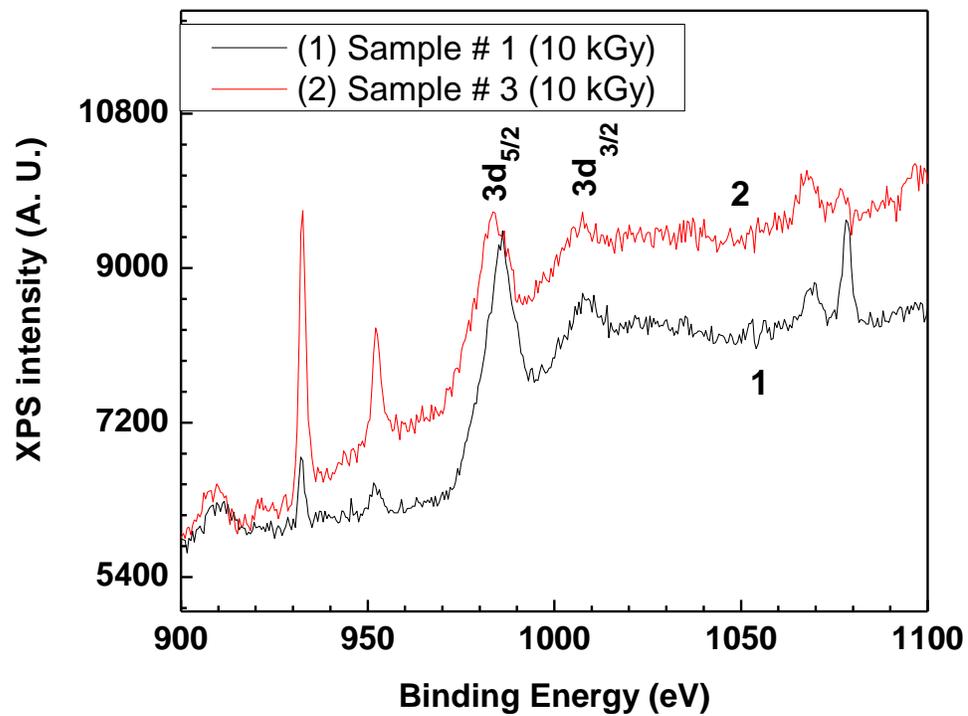

Fig.- 5

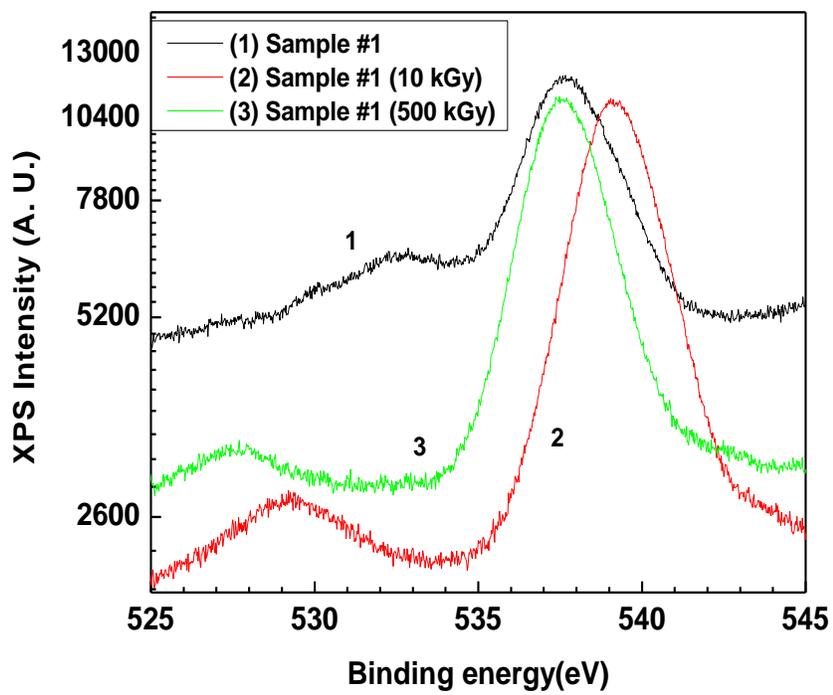

**Fig.- 6**

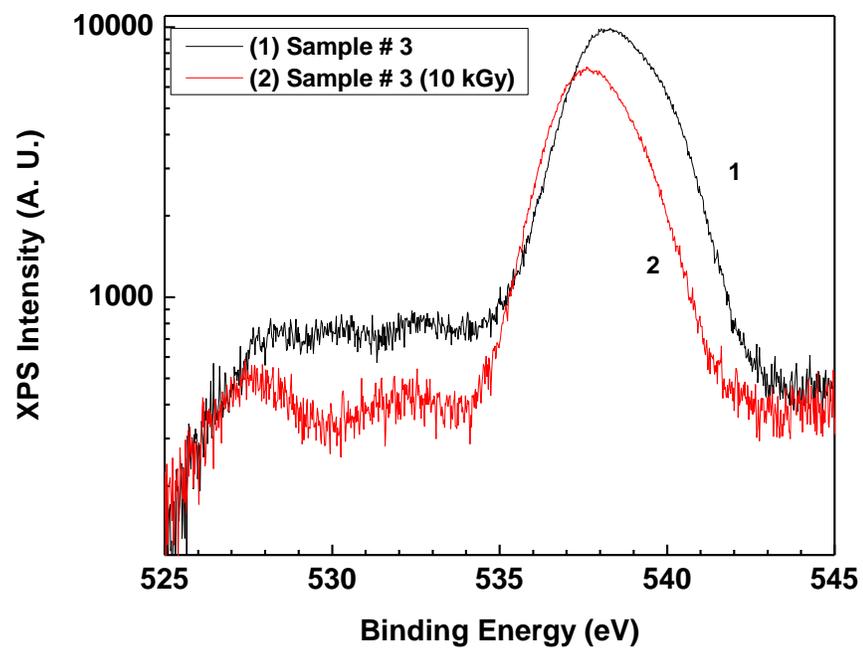

**Fig.-7**